# Engineering the Zero-Point Field and Polarizable Vacuum For Interstellar Flight

H. E. Puthoff[1]
S. R. Little

Institute for Advanced Studies at Austin
4030 W. Braker Ln., Suite 300
Austin, Texas 78759-5329

**ABSTRACT**

A theme that has come to the fore in advanced planning for long-range space exploration is the concept of "propellantless propulsion" or "field propulsion." One version of this concept involves the projected possibility that empty space itself (the quantum vacuum, or space-time metric) might be manipulated so as to provide energy/thrust for future space vehicles.[2] Although such a proposal has a certain science-fiction quality about it, modern theory describes the vacuum as a polarizable medium that sustains energetic quantum fluctuations. Thus the possibility that matter/vacuum interactions might be engineered for space-flight applications is not *a priori* ruled out, although certain constraints need to be acknowledged. The structure and implications of such a far-reaching hypothesis are considered herein.

## 1. INTRODUCTION

The concept of "engineering the vacuum" found its first expression in the mainstream physics literature when it was introduced by T. D. Lee in his textbook *Particle Physics and Introduction to Field Theory*.[3] There he stated: "The experimental method to alter the properties of the vacuum may be called vacuum engineering.... If indeed we are able to alter the vacuum, then we may encounter some new phenomena, totally unexpected." This legitimization of the vacuum engineering concept was based on the recognition that the vacuum is characterized by parameters and structure that leave no doubt that it constitutes an energetic medium in its own right. Foremost among these are its properties that (1) within the context of quantum theory the vacuum is the seat of energetic particle and field fluctuations, and (2) within the context of general relativity the vacuum is the seat of a space-time structure (metric) that encodes the distribution of matter and energy. Indeed, on the flyleaf of a book of essays by Einstein and others on the properties of the vacuum we find the statement "The vacuum is fast emerging as *the* central structure of modern physics."[4]

---

[1] Email: puthoff@earthtech.org
[2] H. E. Puthoff, "Can the vacuum be engineered for spaceflight applications? Overview of theory and experiments," Jour. Sci. Exploration **12**, 295 (1998). See also H. E. Puthoff, "Space propulsion: Can empty space itself provide a solution?" Ad Astra **9** (National Space Society), 42 (Jan/Feb 1997).
[3] T. D. Lee, *Particle Physics and Introduction to Field Theory*, (Harwood Academic Press, London, 1988).

Given the known characteristics of the vacuum, one might reasonably inquire as to why it is not immediately obvious how to catalyze robust interactions of the type sought for space-flight applications. To begin, in the case of quantum fluctuations there are uncertainties that remain to be clarified regarding global thermodynamic and energy constraints. Furthermore, the energetic components of potential utility involve very small-wavelength, high-frequency fields and thus resist facile engineering solutions. With regard to perturbation of the space-time metric, the required energy densities exceed by many orders of magnitude values achievable with existing engineering techniques. Nonetheless, we can examine the constraints, possibilities and implications under the expectation that as technology matures, felicitous means may be found that permit the exploitation of the enormous, as-yet-untapped potential of so-called "empty space."

## 2. PROPELLANTLESS PROPULSION

2.1 Global Constraint

Regardless of the mechanisms that might be entertained with regard to "propellantless" or "field" propulsion of a spaceship, there exist certain constraints that can be easily overlooked but must be taken into consideration. A central one is that, because of the law of conservation of momentum, the center of mass-energy (CM) of an initially stationary isolated system cannot change its position if not acted upon by outside forces. This means that propellantless or field propulsion, whatever form it takes, is constrained to involve coupling to the external universe in such a way that the displacement of the CM of the spaceship is matched by a counteracting effect in the universe to which it is coupled, so as not to violate the global CM constraint. Therefore, before one launches into a detailed investigation of a proposed propulsion mechanism it is instructive to apply this principle as an overall constraint to determine whether the principle is violated. Surprising subtleties may be involved in such an assessment, as illustrated in the following example.

2.2 An Example: "$\mathbf{E} \times \mathbf{H}$" Electromagnetic Field Propulsion

A recurring theme in electromagnetic propulsion considerations is that one might employ crossed electric and magnetic fields to generate propulsive force, what we might call $\mathbf{E} \times \mathbf{H}$ propulsion. The idea is based on the fact that *propagating* electromagnetic fields (photons) possess momentum carried by the crossed (orthogonal) $\mathbf{E}$ and $\mathbf{H}$ fields (Poynting vector). This raises the issue as to whether *static* (i.e., non-propagating) $\mathbf{E} \times \mathbf{H}$ fields also constitute momentum (as the mathematics would imply), and in particular whether changes in static fields could result in the transfer of momentum to an attached structure. As it turns out, the answer can be yes, as illustrated in the example of the

---

[4] *The Philosophy of Vacuum*, Eds. S. Saunders and H. R. Brown (Clarendon Press, Oxford, 1991).

*Feynman disk paradox*.[5]   Electric charge distributed around the rim of a non-rotating disk generates a static electric field that extends outward from the rim, and a current-carrying coil of wire mounted perpendicular to the plane of the disk generates a static dipole magnetic field.  The two fields result in a static $\mathbf{E} \times \mathbf{H}$ distribution that encircles the disk.  Even though nothing is apparently in motion, if we take the $\mathbf{E} \times \mathbf{H}$ momentum concept seriously it would appear that there is angular momentum "circulating" about the disk in the *static* fields.  That this is in fact the case is demonstrated by the fact that when the current in the coil is interrupted, thereby extinguishing the magnetic field component of the $\mathbf{E} \times \mathbf{H}$ distribution, the disk begins to rotate.  This behavior supports the notion that, indeed, the static fields do contain angular momentum that is then transferred to the disk (to conserve angular momentum) when the field momentum is extinguished.[6]  This leads one to wonder if the same principle could be applied to generate linear thrust by changes in static $\mathbf{E} \times \mathbf{H}$ fields, properly arrayed.

Pursuit of the *linear* thrust possibility, however, leads one to a rich literature concerning so-called "hidden momentum" that, perhaps surprisingly, denies this possibility.[7]  The "hidden momentum" phrase refers to the fact that although the linear $\mathbf{E} \times \mathbf{H}$ fields do carry momentum as in the angular case, the symmetry conditions for the linear case are such that there exists a canceling *mechanical* momentum contained in the structures *even though a structure's CM itself is stationary* (see Appendix A). Specifically, it can be shown on very general grounds that, contrary to the case for angular momentum (e.g., the Feynman disk), the total linear momentum of any stationary distribution of matter, charge and their currents, and their associated fields, must vanish. In other words, barring a new discovery that modifies the present laws of physics, any such distribution cannot generate a propulsive force without emitting some form of reaction mass or energy, or otherwise imparting momentum to another system[8].

## 3.  THE QUANTUM VACUUM

### 3.1 Zero-Point Energy (ZPE) Background

Quantum theory tells us that so-called "empty space" is not truly empty, but is the seat of myriad energetic quantum processes.  Specifically, quantum field theory tells us that, even in empty space, fields (e.g., the electromagnetic field) continuously fluctuate about their zero baseline values.  The energy associated with these fluctuations is called zero-point energy (ZPE), reflecting the fact that such activity remains even at a temperature of absolute zero.  Such a concept is almost certain to have profound implications for future space travel, as we will now discuss.

---

[5] R. P. Feynman, R. B. Leighton and M. Sands, *The Feynman Lectures on Physics* (Addison-Wesley, Reading, MA, 1964), Vol. II, p. 17-5.

[6] For experimental confirmation see, for example, G. M. Graham and D. G. Lahoz, "Observation of static electromagnetic angular momentum *in vacuo*," Nature **285**, 154 (1980).

[7] V. Hnizdo, "Hidden momentum of a relativistic fluid carrying current in an external electric field," Am. J. Phys. **65**, 92 (1997).

[8] Proposals to push directly against the space-time metric or quantum vacuum, i.e., use the rest of the Universe as a springboard by means presently unknown, fall into the latter category.

When a hypothetical ZPE-powered spaceship strains against gravity and inertia, there are three elements of the equation that the ZPE technology could in principle address: (1) a decoupling from gravity, (2) a reduction of inertia, or (3) the generation of energy to overcome both.

3.2 Gravity

With regard to a ZPE basis for gravity, the Russian physicist Andrei Sakharov was the first to propose that in a certain sense gravitation is not a fundamental interaction at all, but rather an induced effect brought about by changes in the quantum-fluctuation energy of the vacuum when matter is present.[9] In this view, the attractive gravitational force is more akin to the induced van der Waals and Casimir forces, than to the fundamental Coulomb force. Although quite speculative when first introduced by Sakharov in 1967, this hypothesis has led to a rich literature on quantum-fluctuation-induced gravity. (The latter includes an attempt by one of the authors to flesh out the details of the Sakharov proposal,[10] though difficulties remain.[11]) Given the possibility of a deep connection between gravity and the zero-point fluctuations of the vacuum, it would therefore appear that a potential route to gravity decoupling would be via control of vacuum fluctuations.

3.3 Inertia

Closely related to the ZPE basis for gravity is the possibility of a ZPE basis for inertia. This is not surprising, given the empirical fact that gravitational and inertial masses have the same value, even though the underlying phenomena are quite disparate; one is associated with the gravitational attraction between bodies, while the other is a measure of resistance to acceleration, even far from a gravitational field. Addressing this issue, the author and his colleagues evolved a ZPE model for inertia which developed the concept that although a uniformly moving body does not experience a drag force from the (Lorentz-invariant) vacuum fluctuations, an *accelerated* body meets a resistive force proportional to the acceleration,[12] an approach that has had a favorable reception in the scientific community.[13] Again, as in the gravity case, it would therefore appear that a potential route to the reduction of inertial mass would be via control of vacuum fluctuations.

---

[9] A. D. Sakharov, "Vacuum quantum fluctuations in curved space and the theory of gravitation," Dokl. Akad. Nauk SSSR [Sov. Phys. - Dokl. **12**, 1040 (1968)]. See also C. W. Misner, K. S. Thorne and J. A. Wheeler, *Gravitation*, (Freeman, San Francisco, 1973), pp. 426-428.
[10] H. E. Puthoff, "Gravity as a zero-point-fluctuation force," Phys. Rev. A **39**, 2333 (1989).
[11] H. E. Puthoff, "Reply to 'Comment on "Gravity as a zero-point-fluctuation force,' " Phys. Rev. A **47**, 3454 (1993).
[12] B. Haisch, A. Rueda and H. E. Puthoff, "Inertia as a zero-point field Lorentz force," Phys. Rev. A **49**, 678 (1994); A. Rueda and B. Haisch, "Inertia as reaction of the vacuum to accelerated motion," Phys. Lett. A **240**, 115 (1998).
[13] M. Jammer, *Concepts of Mass in Contemporary Physics and Philosophy*, (Princeton University Press, Princeton, 2000), pp. 163-167.

Investigation into this possibility by the U.S. Air Force's Advanced Concepts Office at Edwards Air Force Base resulted in the generation of a report entitled *Mass Modification Experiment Definition Study* that addressed just this issue.[14] Included in its recommendations was a call for precision measurement of what is called the Casimir force. The Casimir force is an attractive quantum force between closely spaced metal or dielectric plates (or other structures) that derives from partial shielding of the interior region from the background zero-point fluctuations of the vacuum electromagnetic field, which results in unbalanced ZPE radiation pressures.[15] Since issuance of the report, such precision measurements have been made which confirm the Casimir effect to high accuracy,[16] measurements which even attracted high-profile attention in the media.[17] The relevance of the Casimir effect to our considerations is that it constitutes experimental evidence that *vacuum fluctuations can be altered by technological means*. This suggests the possibility that, given the models discussed, *gravitational and inertial masses might also be amenable to modification*. The control of vacuum fluctuations by the use of cavity structures has already found practical application in the field of cavity quantum electrodynamics, where the spontaneous emission rates of atoms are subject to manipulation.[18] Therefore, it is not unreasonable to contemplate the possibility of such control in the field of space propulsion.

3.4 Energy Extraction

With regard to the extraction of energy from the vacuum fluctuation energy reservoir, there are no energetic or thermodynamic constraints preventing such release under certain conditions.[19] And, in fact, there are analyses in the literature that suggest that such mechanisms are already operative in Nature in the "powering up" of cosmic rays,[20] or as the source of energy release from supernovas[21] and gamma-ray bursts.[22]

For our purposes, the question is whether the ZPE can be "mined" at a level practical for use in space propulsion. Given that the ZPE energy density is conservatively estimated to be on the order of nuclear energy densities or greater,[23] it would constitute a seemingly ubiquitous energy supply, a veritable "Holy Grail" energy source.

---

[14] R. L. Forward, "Mass modification experiment definition study," Phillips Laboratory Report PL-TR-96-3004, Air Force Material Command, Edwards AFB, CA 93524-5000 (Feb 1996). Reprinted in Jour. Sci. Exploration **10**, 325 (1996).

[15] P. W. Milonni, R. J. Cook and M. E. Goggin, "Radiation pressure from the vacuum: Physical interpretation of the Casimir force," Phys. Rev. A **38**, 1621 (1988).

[16] S. K. Lamoreaux, "Demonstration of the Casimir force in the 0.6 to 6 μm range," Phys. Rev. Lett. **78**, 5 (1997); U. Mohideen and A. Roy, "Precision measurement of the Casimir force from 0.1 to 0.9 μm," Phys. Rev. Lett. **81**, 4549 (1998).

[17] M. W. Browne, "Physicists confirm power of nothing, measuring force of quantum foam," New York Times, p. C1, 21 Jan. 1997.

[18] See, e.g., S. Haroche and J.-M. Raimond, "Cavity quantum electrodynamics," Sci. Am. p. 54 (April 1993).

[19] D. C. Cole and H. E. Puthoff, "Extracting energy and heat from the vacuum," Phys. Rev. E **48**, 1562 (1993).

[20] A. Rueda, B. Haisch and D. C. Cole, "Vacuum zero-point field pressure instability in astrophysical plasmas and the formation of cosmic voids," Astrophys. J. **445**, 7 (1995).

[21] I. Tu. Sokolov, "The Casimir effect as a possible source of cosmic energy," Phys. Lett. A **223**, 163 (1996).

[22] C. E. Carlson, T. Goldman and J. Peres-Mercader, "Gamma-ray bursts, neutron star quakes, and the Casimir effect," Europhys. Lett. **36**, 637 (1996).

One of the first researchers to call attention to the principle of the use of the Casimir effect as a potential energy source was Robert Forward at Hughes Research Laboratories in Malibu, CA.[24] Though providing "proof-of-principle," unlike the astrophysical implications cited above the amount of energy release for mechanical structures under laboratory conditions is minuscule. In addition, the conservative nature of the Casimir effect would appear to prevent recycling, though there have been some suggestions for getting around this barrier.[25] Alternatives involving non-recycling behavior, such as plasma pinches[26] or bubble collapse in sonoluminescence,[27] have been investigated in our laboratory and elsewhere, but as yet without real promise for energy applications.

Vacuum energy extraction approaches by other than the Casimir effect are also being considered. One approach that emerged from the Air Force's *Mass Modification…* study (Ref. 14) was the suggestion that the ZPE-driven cosmic ray model be explored under laboratory conditions to determine whether protons could be accelerated by the proposed cosmic ray mechanism in a cryogenically-cooled, collision-free vacuum trap. Yet another proposal (for which a patent has been issued) is based on the concept of beat-frequency downshifting of the more energetic high-frequency components of the ZPE, by use of slightly detuned dielectric-sphere antennas.[28]

In our own laboratory we have considered an approach based on perturbation of atomic or molecular ground states, hypothesized to be equilibrium states involving dynamic radiation/absorption exchange with the vacuum fluctuations.[29] In this model atoms or molecules in a ZPE-limiting Casimir cavity are expected to undergo energy shifts that would alter the spectroscopic signatures of excitations involving the ground state. We have initiated experiments at a synchrotron facility to explore this ZPE/ground-state relationship, though so far without success.

---

[23] R. P. Feynman and A. R. Hibbs, *Quantum Mechanics and Path Integrals*, (McGraw-Hill, New York, 1965).

[24] R. L. Forward, "Extracting electrical energy from the vacuum by cohesion of charged foliated conductors," Phys. Rev. B **30**, 1700 (1984).

[25] F. Pinto, "Engine cycle of an optically controlled vacuum energy transducer," Phys. Rev. B **60**, 14, 740 (1999).

[26] H. E. Puthoff, "The energetic vacuum: implications for energy research," Spec. in Sci. and Tech. **13**, 247 (1990); K. R. Shoulders, "Method and apparatus for production and manipulation of high density charge," U.S. Patent No. 5,054,046 (1991).

[27] J. Schwinger, "Casimir light: the source," Proc. Nat'l Acad. Sci. **90**, 2105 (1993); C. Eberlein, "Sonoluminescence as quantum vacuum radiation," Phys. Rev. Lett. **76**, 3842 (1996).

[28] F. B. Mead and J. Nachamkin, "System for converting electromagnetic radiation energy to electrical energy," U.S. Patent No. 5,590,031 (1996).

[29] H. E. Puthoff, "Ground state of hydrogen as a zero-point-fluctuation-determined state," Phys. Rev. D **35**, 3266 (1987).

Whether tapping the ZPE as an energy source or manipulating the ZPE for gravity/inertia control are but gleams in a spaceship designer's eye, or a Royal Road to practical space propulsion, is yet to be determined. Only by explorations of the type described here will the answer emerge. In the interim a quote by the Russian science historian Roman Podolny would seem to apply: "It would be just as presumptuous to deny the feasibility of useful application as it would be irresponsible to guarantee such application."[30]

## 4. THE SPACE-TIME METRIC ("METRIC ENGINEERING" APPROACH)

Despite the apparently daunting energy requirements to perturb the space-time metric to a significant degree, we examine the structure that such perturbations would take under conditions useful for space-flight application, a "Blue Sky" approach, as it were.

Although topics in general relativity are routinely treated in terms of tensor formulations in curved space-time, we shall find it convenient for our purposes to utilize one of the alternative methodologies for treating metric changes that has emerged over the years in studies of gravitational theories. The approach, known as the polarizable vacuum (PV) representation of general relativity (GR), treats the vacuum as a polarizable medium.[31] The PV approach treats metric changes in terms of the permittivity and permeability constants of the vacuum, $\varepsilon_o$ and $\mu_o$, essentially along the lines of the "$TH\varepsilon\mu$" methodology used in comparative studies of gravitational theories.[32] Such an approach, relying as it does on parameters familiar to engineers, can be considered a "metric engineering" approach.

In brief, Maxwell's equations in curved space are treated in the isomorphism of a polarizable medium of variable refractive index in flat space;[33] the bending of a light ray near a massive body is modeled as due to an induced spatial variation in the refractive index of the vacuum near the body; the reduction in the velocity of light in a gravitational potential is represented by an effective increase in the refractive index of the vacuum, and so forth. As elaborated in Ref. 31 and the references therein, though differing in some aspects from GR, PV modeling can be carried out for cases of interest in a self-consistent way so as to reproduce to appropriate order both the equations of GR, and the match to the classical experimental tests of those equations.

---

[30] R. Podolny, *Something Called Nothing - Physical Vacuum: What is It?*, (Mir Publishers, Moscow, 1986).
[31] H. E. Puthoff, "Polarizable-vacuum approach to general relativity," in *Gravitation and Cosmology: From the Hubble Radius to the Planck Scale*, Eds. R. L. Amoroso, G. Hunter, M. Kafatos, and J.-P. Vigier (Kluwer Academic Press, Dordrecht, the Netherlands, in press, 2001). See also H. E. Puthoff, "Polarizable-vacuum (PV) representation of general relativity," http://xxx.lanl.gov/abs/gr-qc/9909037.
[32] A. P. Lightman and D. L. Lee, "Restricted proof that the weak equivalence principle implies the Einstein equivalence principle," Phys. Rev. D **8**, 364 (1973).
[33] A. M. Volkov, A. A. Izmest'ev and G. V. Skrotskii, "The propagation of electromagnetic waves in a Riemannian space," Sov. Phys. JETP **32**, 686 (1971).

Specifically, the PV approach treats such measures as the velocity of light, the length of rulers (atomic bond lengths), the frequency of clocks, particle masses, and so forth, in terms of a variable vacuum dielectric constant $K$ in which vacuum permittivity $\varepsilon_o$ transforms to $\varepsilon_o \to K\varepsilon_o$, vacuum permeability to $\mu_o \to K\mu_o$. In a planetary or solar gravitational potential $K \approx 1 + 2GM/rc^2 > 1$, and the results are as shown in Table 1. Thus, the velocity of light is reduced, light emitted from an atom is redshifted as compared with an atom at infinity ($K = 1$), rulers shrink, etc.

As one example of the significance of the tabulated values, the dependence of fundamental length measures (ruler shrinkage) on the variable $K$ indicates that the dimensions of material objects adjust in accordance with local changes in vacuum polarizability - - thus there is no such thing as a perfectly rigid rod. From the standpoint of the PV approach this is the genesis of the variable metric that is of such significance in GR studies. It also permits us to define, from the viewpoint of the PV approach, just what precisely is meant by the label "curved space." In the vicinity of, say, a planet or star, where $K > 1$, if one were to take a ruler and measure along a radius vector $R$ to some circular orbit, and then measure the circumference $C$ of that orbit, one would obtain $C < 2\pi R$ (as for a concave curved surface). This is a consequence of the ruler being relatively shorter during the radial measuring process when closer to the body where $K$ is relatively greater, as compared to its length during the circumferential measuring process when further from the body. Such an influence on the measuring process due to induced polarizability changes in the vacuum near the body leads to the GR concept that the presence of the body "influences the metric," and correctly so.

**Table 1**
**Typical Metric Effects in the Polarizable Vacuum (PV) Representation of GR**

(For reference frame at infinity, $K = 1$)

| *Variable* | *Determining Equation* | *K≥1* (*typical mass distribution, M*) |
|---|---|---|
| velocity of light $v_L(K)$ | $v_L = c/K$ | velocity of light $< c$ |
| mass $m(K)$ | $m = m_o K^{3/2}$ | effective mass increases |
| frequency $\omega(K)$ | $\omega = \omega_o/\sqrt{K}$ | redshift toward lower frequencies |
| time interval $\Delta t(K)$ | $\Delta t = \Delta t_o \sqrt{K}$ | clocks run slower |
| energy $E(K)$ | $E = E_o/\sqrt{K}$ | lower energy states |
| length dim. $L(K)$ | $L = L_o/\sqrt{K}$ | objects shrink |

We are now in a position to consider application of this "metric engineering" formalism to the type of questions relevant to space propulsion. As we show in Appendix B, under certain conditions the metric can in principle be modified to reduce the value of the vacuum dielectric constant $K$ to below unity. Returning to Table 1, we see that a $K < 1$ solution permits the addition of another column for which the descriptors are reversed, as shown in Table 2.

**Table 2**
**Engineered Metric Effects in the Polarizable Vacuum (PV) Representation of GR**

(For reference frame at infinity, $K = 1$)

| *Variable* | *Determining Equation* | *K≤1 (engineered metric)* |
|---|---|---|
| velocity of light $v_L(K)$ | $v_L = c/K$ | velocity of light $> c$ |
| mass $m(K)$ | $m = m_o K^{3/2}$ | effective mass decreases |
| frequency $\omega(K)$ | $\omega = \omega_o/\sqrt{K}$ | blueshift toward higher frequencies |
| time interval $\Delta t(K)$ | $\Delta t = \Delta t_o \sqrt{K}$ | clocks run faster |
| energy $E(K)$ | $E = E_o/\sqrt{K}$ | higher energy states |
| length dim. $L(K)$ | $L = L_o/\sqrt{K}$ | objects expand |

Under such conditions of extreme space-time perturbation, the *local velocity of light* (as seen from a reference frame at infinity) *is increased, mass decreases, energy bond strengths increase*, etc., features presumably attractive for interstellar travel.[34]

Clearly, calculations for such simple geometries are by no means directly applicable to the design of a space propulsion drive. However, these sample calculations indicate the direction of potentially useful trends derivable on the basis of the application of GR principles as embodied in the metric engineering approach, with the results constrained only by what is achievable practically in an engineering sense.

## 5. CONCLUSIONS

In this paper we have touched briefly on innovative forms of space propulsion, especially those that might exploit properties of the quantum vacuum or the space-time metric in a fundamental way. At this point in the development of such nascent concepts it is premature to even guess at an optimum strategy, let alone attempt to forge a critical path; in fact, it remains to be determined whether such exploitation is even feasible. Nonetheless, only by inquiring into such concepts in a rigorous way can we hope to arrive at a proper assessment of the possibilities and thereby determine the best course of action to pursue in our steps first to explore our solar system environment, and then one day to reach the stars.

---

[34] See, for example, H. E. Puthoff, "SETI, the velocity-of-light limitation, and the Alcubierre warp drive: An integrating overview," Phys. Essays **9**, 156 (1996), and the references therein.

## APPENDIX A
## HIDDEN MOMENTUM

Consider a stationary current loop which consists of an incompressible fluid of positive charge density $\rho$ circulating at velocity $v$ clockwise around a loop of non-conductive piping of cross sectional area $a$. The loop is immersed in a constant uniform electric field **E**.

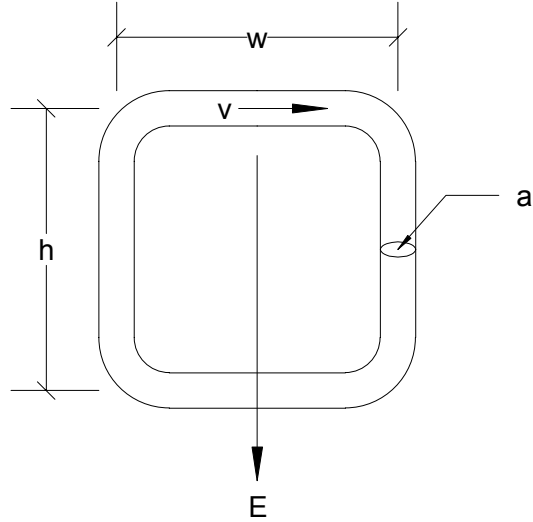

The magnetic field created by the current loop combines with the electric field to produce an electromagnetic field momentum given by

$$\mathbf{p}_{EM} = \frac{1}{c^2} \int \mathbf{E} \times \mathbf{H} \, dV. \tag{A1}$$

However, in steady state situations, this is equal to (Ref. 7)

$$\mathbf{p}_{EM} = \frac{1}{c^2} \int \mathbf{J} \phi \, dV. \tag{A2}$$

With reference to the above figure, the only non-zero component of momentum surviving this integration is directed horizontally across the page. Using the expression Eq. (A2), this computes to

$$p_{EM} = \frac{Iw}{c^2}\left(\phi_{top} - \phi_{bottom}\right) = \frac{IEhw}{c^2}, \tag{A3}$$

where the sense is from left to right. From this it is concluded that there is a steady net *linear* momentum stored in the electromagnetic fields. We will now show there is another momentum, equal and opposite to this electromagnetic field momentum.

Since the current flowing in the loop is given by $I = \rho a v$, the velocity of the fluid is everywhere $v = I/\rho a$. Meanwhile, the external electric field $E$ creates a pressure difference between the bottom and the top of the fluid given by $P = \rho E h$. Moving to the *left*, therefore, is a net energy flux *S* (energy per unit area per unit time) given by

$$S = Pv = (\rho E h) \times (I/\rho a) = \frac{IEh}{a}. \tag{A4}$$

But since energy has mass, Eq. (A4) may be converted to an expression for momentum. This is mostly easily accomplished by writing the Einstein relation $E = mc^2$ in flux density form as $S = gc^2$, where *g* is the momentum per unit volume. It now follows that, due to the different pressures at the top and bottom of the loop, there must be a net overall momentum - *directed to the left* - given by

$$p_{mech} = gaw = \frac{Saw}{c^2} = \frac{IEhw}{c^2}, \tag{A5}$$

where the subscript 'mech' draws attention to the apparently entirely mechanical origin of this momentum.

Eqs. (A5) and (A3) demonstrate that the electromagnetic momentum is balanced by an equal and opposite mechanical momentum. Because of its rather obscure nature, this momentum has been referred to in the literature as "hidden momentum". This is a particular example of the general result that a net static linear field momentum will always be balanced by an equal and opposite hidden mechanical momentum. In practical terms, this means that the creation of linear field momentum cannot give rise to motion because the field momentum is automatically neutralized by a mechanical momentum hidden within the structure, so that the whole system remains stationary. This inability to utilize linear field momentum for propulsion is guaranteed by the law of momentum conservation.

## APPENDIX B
## METRIC ENGINEERING SOLUTIONS

In the polarizable vacuum (PV) approach the equation that plays the role of the Einstein equation (curvature driven by the mass-energy stress tensor) is (Ref. 31)

$$\nabla^2 \sqrt{K} - \frac{1}{(c/K)^2} \frac{\partial^2 \sqrt{K}}{\partial t^2} = -\frac{\sqrt{K}}{4\lambda} \left\{ \frac{(m_0 c^2/\sqrt{K}) \left[1 + \left(\frac{v}{c/K}\right)^2\right]}{\sqrt{1 - \left(\frac{v}{c/K}\right)^2}} \left[\frac{1 + \left(\frac{v}{c/K}\right)^2}{2}\right] \delta^3(\mathbf{r} - \bar{\mathbf{r}}) \right.$$

$$\left. + \frac{1}{2}\left(\frac{B^2}{K\mu_O} + K\varepsilon_O E^2\right) - \frac{\lambda}{K^2}\left[(\nabla K)^2 + \frac{1}{(c/K)^2}\left(\frac{\partial K}{\partial t}\right)^2\right] \right\} \tag{B1}$$

In this PV formulation of GR, changes in the vacuum dielectric constant *K* are driven by mass density (first term), EM energy density (second term), and the vacuum polarization energy density itself (third term). (The constant $\lambda = c^4/32\pi G$, where *G* is the gravitational constant.)

In space surrounding an uncharged spherical mass distribution (e.g., a planet) the static solution *(∂K/∂t = 0)* to the above is found by solving

$$\frac{d^2\sqrt{K}}{dr^2} + \frac{2}{r}\frac{d\sqrt{K}}{dr} = \frac{1}{\sqrt{K}}\left(\frac{d\sqrt{K}}{dr}\right)^2. \tag{B2}$$

The solution that satisfies the Newtonian limit is given by

$$K = \left(\sqrt{K}\right)^2 = e^{2GM/rc^2} = 1 + 2\left(\frac{GM}{rc^2}\right) + ..., \tag{B3}$$

which can be shown to reproduce to appropriate order the standard GR Schwarzschild metric properties as they apply to the weak-field conditions prevailing in the solar system.

For the case of a mass $M$ with charge $Q$, the electric field appropriate to a charged mass imbedded in a variable-dielectric-constant medium is given by

$$\int \mathbf{D} \cdot d\mathbf{a} = K\varepsilon_O E 4\pi r^2 = Q, \tag{B4}$$

which leads to (for spherical symmetry, with $b^2 = Q^2 G / 4\pi\varepsilon_o c^4$)

$$\frac{d^2\sqrt{K}}{dr^2} + \frac{2}{r}\frac{d\sqrt{K}}{dr} = \frac{1}{\sqrt{K}}\left[\left(\frac{d\sqrt{K}}{dr}\right)^2 - \frac{b^2}{r^4}\right], \tag{B5}$$

which should be compared with Eq. (B2). The solution here as a function of charge (represented by $b$) and mass (represented by $a = GM/c^2$) is given by

$$\sqrt{K} = \cosh\left(\frac{\sqrt{a^2 - b^2}}{r}\right) + \frac{a}{\sqrt{a^2 - b^2}}\sinh\left(\frac{\sqrt{a^2 - b^2}}{r}\right), \quad a^2 > b^2. \tag{B6}$$

For the weak-field case the above reproduces the familiar Reissner-Nordstrøm metric.[35] For $b^2 > a^2$, however, the hyperbolic solutions turn trigonometric, and $K$ can take on values $K < 1$.

---

[35] C. W. Misner, K. S. Thorne, and J. A. Wheeler, *Gravitation*, (Freeman, San Francisco, 1973), p. 841.